\documentclass[twocolumn]{revtex4-1}
\UseRawInputEncoding
\usepackage{graphicx}
\usepackage{dcolumn}
\usepackage{bm}
\usepackage{natbib}
\usepackage{xcolor}
\usepackage{verbatim}

\renewcommand{\selectlanguage}[1]{}

\begin{document}

\title{Motion of a floating sphere pulled by a string and induced flows}
\author{Benjamin Apffel}
 \email{benjamin.apffel@epfl.ch}
\author{Romain Fleury}%
\affiliation{%
Institute of Electrical and Micro Engineering, Laboratory of Wave Engineering, Ecole Polytechnique Federale de Lausanne (EPFL), Station 11, 1015 Lausanne, Switzerland
}%

\date{\today}
\begin{abstract}
We study in this article the motion of a floating ball attached to
a soft string set in circular motion through its other end. Although simple, the system exhibits rich dynamics that we investigate experimentally and theoretically.   At low rotation speeds, we show that the circular trajectory of the ball shrinks when we stir faster, which challenges common intuition based on centrifugal force. For higher rotation rates, the ball is either suddenly attracted toward the center, or is repulsed away from it, depending on the string length. Experimental measurements of the generated flow show that a Magnus force must be taken into account to correctly explain all the observations. In particular, our deformable system allows to measure the ratio of the lift force over the inertial force. Interestingly, the system exhibits strong hysteretic behavior, showing that the ball can robustly trap itself at the center of the flow generated by its past motion. The present experiment also revisits the famous 'tea-leaf paradox', which refers to the unexpected migration of tea leaves toward the center of a tea cup when the latter is mixed with a rigid spoon. The ball attached to the string plays the role of a deformable spoon, and we show that there exists a maximum rotation speed above which the tea-leaves transport brutally stops.
\end{abstract}

\maketitle

\section{Introduction}
In classical mechanics, a circular trajectory is associated with a fictitious force that tends to push objects away from the rotation center, in accordance with our sensitive experience during a turn in a ride. However, this intuition is challenged by other experiments in which the reverse effect appears. For instance, particles suspended in a rotating fluid are pushed toward the rotation center if their relative density is less than one. The centrifugal force acts in this case as an effective radial gravity which generates buoyancy directed toward the rotation center for light particles. More surprisingly, heavy particles placed at the bottom of a liquid stirred from the top by a rotating paddle also tend to aggregate at the rotation center. This effect is often refereed to as 'tea-leaf paradox', as it can be observed with tea leaves when one mixes a tea cup with a spoon. Such peculiar particles transport finds its roots in the three-dimensional flows induced by the container edges \cite{thomson_v_1997} and has manifestation in sediments transport in  rivers \citep{einstein_ursache_1926,friedkin_laboratory_1945,bowker_albert_2007, leopold_river_1960,jackson_velocitybed-formtexture_1975}, dissolution mediated by rotating paddles \cite{mccarthy_simulating_2003}, aggregation control of nanoparticles \cite{zhang_einsteins_2023} or blood-plasma separation \citep{yeo_electric_2006,arifin_microfluidic_2006}.

However, the picture is completely different if the stirrer's trajectory can be affected in return by the flow it generates. In the experiments mentioned above, the circular trajectory of the stirrer is indeed imposed by an external operator, and the backaction of the flow simply consists in a friction force that the operator must overcome. The situation is completely different when the trajectory itself can be modified by the action of the flow, a commonly encountered example being the deflection experienced by a spinning sphere moving in a fluid due to lift (or Magnus) forces \cite{clanet_sports_2015,dupeux_spinning_2010,rubinow_transverse_1961}. Similarly, the trajectory of magnetic disks set in rotation at a liquid interface were shown to be strongly impacted by the generated flow \citep{gorce_confinement_2019,gorce_rolling_2021}, and several disks could interact and self-assemble under the generated flow's action \citep{grzybowski_dynamic_2000,grzybowski_dynamic_2001,grzybowski_dynamics_2002}. On top of those translational degrees of freedom, object-flow interactions can also alter the shape of deformable bodies. For instance, a flexible plate placed in constant flow will tend to bend, which will modify the flow in return\citep{wu_swimming_1961, alben_collective_2021,toomey_numerical_2008}. More generally, the interplay between deformable bodies and flow generation is at the heart of propulsion mechanisms for living organisms \citep{gazzola_scaling_2014,taylor_analysis_1997,wu_fish_2011,ashraf_simple_2017} or robots \citep{oliveira_santos_pleobot_2023,esposito_robotic_2012,zhu_tuna_2019}. The extra degrees of freedom offered by the object’s deformations and motion therefore offer a large panel of behaviors that are still to explore.

Inspired by this rich phenomenology, we propose to revisit the tea-leaf paradox when one replaces the rigid spoon by a deformable one, as depicted in Fig. \ref{fig:0}a. As an experimental model of such system, we will study the motion of a ball attached to a rotation arm by a deformable string as (Fig. \ref{fig:0}b). The motion at one end of the string is therefore imposed, but the ball attached to other end can move under the liquid's action. Although simple in principle, this system exhibits rich phenomenology that we briefly summarize here.  At low rotation speed, the ball performs circular trajectories which radius decreases when the rotation rate increases, hence challenging the intuition on centrifugal force. At high rotation speed, a strong bifurcation is observed and the ball suddenly gets  expelled or attracted toward the rotation center depending on the string length. In the latter case, the ball can moreover exhibit strong hysteretic behavior and gets robustly trap at the rotation center, showing that the system is sensitive to its past history. Moreover, those states are shown to drastically reduce the mixing efficiency of the liquid.

This work is organized as follow. We first present in section II experimental results for different rotation rates and string length. Section III is dedicated to force analysis while section IV focuses on the lift force exerted by the liquid, which must be taken into account to satisfactory explain the experimental results. Section V focuses on the impact of string deformation, and a full model of the system which shows good agreement with our data is proposed. We discuss in section VI the appearance of hysteresis and discuss the existence of self-trapped states. Section VII finally discusses the emergence of tea-leaf paradox in our experimental case.

\section{Experimental setup and results}

\begin{figure}
    \centering
    \includegraphics[width=8.5cm]{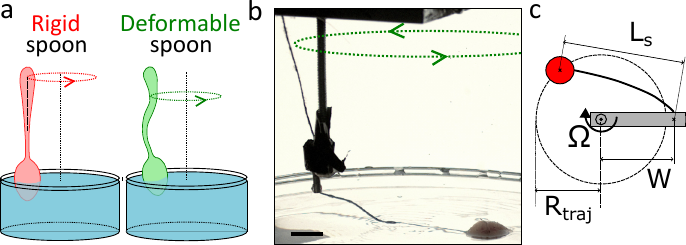}
    \caption{Experimental system and definitions. (a) What happens if one tries to stirr a liquid with a deformable spoon ? (b) Picture of the model deformable spoon consisting in a ball pulled by an inextensible string. The latter is guided by a tube and knotted to the rotation arm on the top. (c) Definition of the string length $L_s$ and trajectory radius $R_t$, rotation arm length being fixed to $W=4.4$ cm.}
    \label{fig:0}
\end{figure}

\begin{figure*}
    \centering
    \includegraphics[width=17cm]{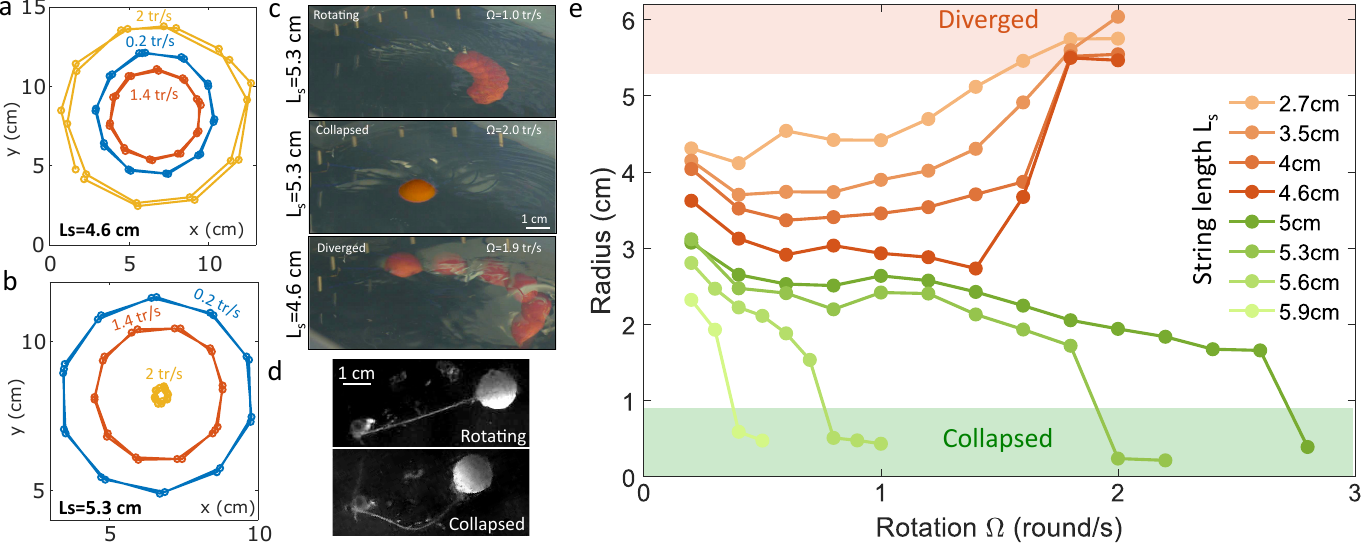}
    \caption{Trajectory of the ball for different string lengths when the rotation speed is progressively increased. (a-b) Examples of experimentally measured trajectories for different rotation speed $\Omega=0.2, 1.4, 2$ round/s and two different string length (a) $L_s = 4.6$ cm and (b) $L_s = 5.3$ cm. (c) Experimental kinogrammes showing a circular (top), a collapsed  (middle) and a diverged (bottom) trajectory. (d) Picture from below showing the loss of tension in the string when a collapsed state is reached. (e) Radius of the trajectory for progressively increasing rotation speed and different string lengths $L_s$. For the shortest strings ($L_s \leq 4.6$ cm, red curves), the ball reaches a diverged state while longest strings ($L_s \geq 5$ cm, green curves) lead to collapsed states.}
    \label{fig:1}
\end{figure*}

We describe in this section our experimental setup and results. All measurement are conducted in a rectangular tank of 26$\times$36 cm$^2$ filled with tap water with depth 6.5 cm. Our model deformable stirrer consists in a 3D-printed plastic ball of radius $R_s=7$~mm, mass $m=1.3$~g and relative density $\hat \rho_s = 0.9$ attached to a rotation arm of length $W=4.4$~cm by a wool strand, as shown in Fig. \ref{fig:0}b-c. The arm is rotated at rate $\Omega$ going from 0.2 to 3 round/s with a stepper motor. In all the experiments, the ball remains far ($>3$ cm) from any edge of the container to prevent boundary effect to occur. In order to recover the ball's trajectory, a camera (Basler) is placed below the tank and records 10 images per revolution of the rotation arm. The ball's position is retrieved on each image using a convolution algorithm. More details on the experimental procedure and analysis algorithm can be found in supplementary materials. 

In our experimental regimes, the strand is non-extensible, of negligible mass compared to the ball and free of static or plastic deformation after being constrained. Nevertheless, the way to attach it to the rotation arm is a sensitive parameter for experimental repeatability. After several attempts, the final design consists in knotting the strand to the rotation arm and guide it close to the surface with a tube as shown in Fig. \ref{fig:0}b. This avoids the introduction of constrains, as the strand is free to move in the tube, and confines the deformable part of the stirrer close to the horizontal interface. In this configuration the camera placed below the tank can also be used to characterize the string's shape, as most of its deformation will occur in the horizontal plane. The results obtained with this setup were qualitatively similar to those obtained when the strand was directly tied to the rotation arm far from the surface. This shows that confining the strand in the horizontal plane did not affect significantly the underlying physics while simplifying experimental measurement and enforcing repeatability.   

We performed experimental measurement of the ball's trajectory for different rotation speeds $\Omega$ and different string lengths $L_s$, measured from the tube to the ball. We plot in Fig. \ref{fig:1}a-b the results for three different rotation speed $\Omega=0.2, 1.4$ and $2.0$ round/s and two string length $L_s = 4.6$ cm (Fig. \ref{fig:1}a) and $L_s = 5.3$ cm (Fig. \ref{fig:1}b). The trajectories are circular in all cases and their radius $R_t$ depends strongly on the rotation speed. For both string length, the radius of the trajectory decreases as $\Omega$ goes from $0.2$ to $1.4$ round/s. This is opposite to what is expected from the tea-leaf paradox, in which floating particles are pushed toward the outside by secondary flows. When the rotation speed is increased to $2.0$ round/s, two opposite behavior were observed depending on the string length. For the shortest string, the ball was suddenly expelled from the rotation center in what we call a 'diverged' state. Conversely, the ball was trapped at the rotation center in a 'collapsed' state when the string was longer. Kinnograms obtained by stacking images from the side are shown in Fig. \ref{fig:1}c and display the three 'rotating', 'collapsed' and 'diverged' regimes discussed above. Experimental snapshots in Fig. \ref{fig:1}d show that the string is straight in non-collapsed cases, but gains some curvature in the collapsed state. One can therefore qualitatively interpret such collapse as the inability of the system to maintain tension in the string during the ball's straddling

More systematic measurement of the trajectory's radius $R_t$  for different rotation rate $\Omega$ and string length $L_s$ are shown in Fig. \ref{fig:1}e. Two distinct behaviors emerge from those results. For the shortest string length ($L_s<5$ cm, red curves), the ball eventually reaches a diverged state at high rotation speed, where the radius no longer depends on the string length. This state is reached continuously for the shortest string length ($L_s = 2.7$ cm and $3.5$ cm) or by a brutal jump for the longest strings ($L_s =4$ cm and $4.6$ cm). Before such diverged state is reached, there exists a range in which the trajectory radius decreases when the rotation speed increases. The associated range of rotation speeds increases as the string lengthens. Above a certain string length ($L_s \geq 5$ cm, green curves), the ball conversely reaches a collapsed state. The rotation speed at which the collapse occurs strongly decreases as the string becomes longer, going from $2.8$ round/s for $L_s = 5$ cm to almost zero for $L_s > 6$ cm. Before such collapse occurs, we observe as for shorter string length that the radius globally decreases when the rotation speed increases. 

In what follows, we aim to explain theoretically both why the radius decreases when the rotation speed increases and why the system collapses or diverges depending on the string length.

\section{Force analysis}

\begin{figure*}
    \centering
    \includegraphics[width=17cm]{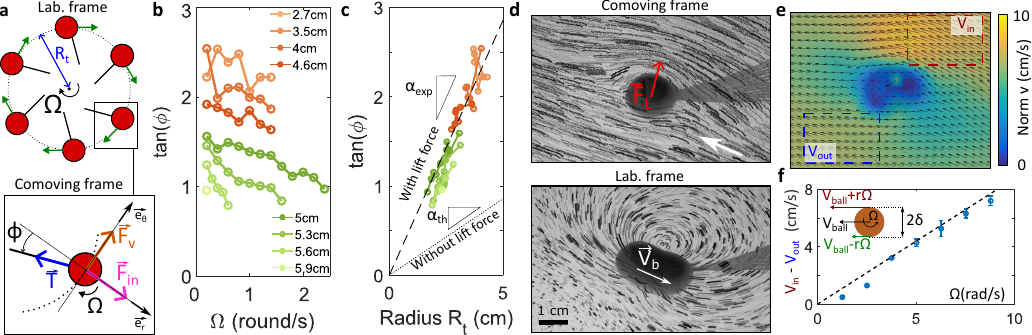}
    \caption{Force analysis. (a) Sketch of the ball trajectory in the laboratory or comoving frame, and definition of the forces and angle $\phi=( -\vec e_r, \vec T)$. (b) Experimental measurement of $\tan(\phi)$ for different string length as a function of rotation speed $\Omega$. (c) All data collapse in on single line when plotted as a function of the trajectory radius $R_t$. Its slope is $\alpha_{exp}=0.56$ cm$^{-1}$ (dashed line) while theory without lift force predicts $\alpha_{th} = 0.17$ cm$^{-1}$ (dot line). Adding the lift force allows to correct the slope to match experimental. (d) Stack of 15 consecutive images taken at 150 frame/s showing the motion of particles at the water surface in the laboratory (left) and comoving (right) frame. Fluid motion (white arrow) is indicated and according to the streamlines and an extra lift force $\vec F_L$ (red arrow) is expected to occur. (e) Example of velocity reconstruction using PIV (black arrows) and corresponding norm (background color). The fluid moves faster on the inner side (red square) than on the outer side (blue square) of the sphere. (f) Averaged velocity difference $V_{in} - V_{out}$ for different rotation speeds. The errorbars show the standard deviation of the measurement over 20 consecutive frames.}
    \label{fig:2}
\end{figure*}

We propose in this section to perform a force analysis on the ball in the comoving frame. For now, we will consider the inertial force $\vec F_{in}$, the viscous force $\vec F_v$ and the string tension $\vec T$ as sketched in Fig. \ref{fig:2}a. We assume that the string is tensed and that the ball performs a circular motion of radius $R_t$ with rotation speed $\Omega$.The inertial force due to the acceleration writes $\vec F_{in} =  (m + m_{add} )  R_t \Omega^2 \vec e_r$
where $m_{add}$ is the added mass of fluid that needs to be pushed by the sphere. In the potential flow approximation, the added mass of a fully immersed sphere can be computed to be exactly $m_{add}=m/2$ \cite{falkovich_fluid_2018}. However, the flow is not irrotational (see Fig. \ref{fig:2}d) and the presence of a free surface makes the computation of the added mass a very difficult task. In particular, the waves emitted during the ball's motion may significantly contribute to the added mass \cite{ursell_heaving_1949,voisin_added_2024} We have measured experimentally that the waves emitted had a maximum amplitude of typically $A\sim 0.2$ mm with wavevector $k \sim 2\pi/R_s$ (see Supplementary Materials - SM). A plane wave of such amplitude would apply on an object of size $R_s$ a force of typically $F_{wave} \sim 1/4 (\rho g + 3\gamma k^2) A^2 R_s$ \cite{longuet-higgins_radiation_1964}, and one can thus show that $F_{wave}/ F_{in} < 10^{-2}$. This justifies that we neglect the impact of the waves in what follows. Nevertheless the free surface changes the boundary condition compared to infinite fluid, which impacts the exact value of the added mass. In what follows, we will thus write $m_{add} = \hat M _{add} m$ and keep $\hat M_{add} \sim 1/2$ as an approximated value for discussion.

The Reynolds number $Re$ typically ranges from $500$ to $5000$ in our experiments, so that a quadratic law can be used to estimate the friction force $\vec F_v = 1/2 \rho C_D \hat S_{im} \pi R_s^2 (R_t \Omega) ^2 \vec e_\theta$, where $C_D$ is the dimensionless drag parameter \cite{clanet_sports_2015} and $\hat S_{im}$ is the immersed cross-sectional surface of the object in the direction of motion normalized by the total normal surface $\pi R_s^2$. For simplicity, we will assume that the immersed volume does not vary with the rotation speed. Experimental pictures taken from the side confirm this hypothesis except in the diverged regime, in which the immersed volume diminishes significantly (see SM). The corresponding data will hence be discarded for the force analysis. We will furthermore only consider the cases were the string is tensed, and therefore discard all the collapsed states as well. An extensive discussion on the data selection is available in SI. 

As the ball is steady in the comoving frame, one obtains $\vec T = -\vec F_{in} - \vec F_{v}$. In particular, the angle $\phi = (-\vec e_r, \vec T)$ drawn in Fig.\ref{fig:2}a can be expressed as
\begin{equation}
    \tan(\phi)= \frac{F_v}{F_{in}} = \alpha_{th} R_t
    \label{eq:phiD}
\end{equation}
where 
\begin{equation}
    \alpha_{th}= \frac{3C_D \hat S_{im}}{8 (1 + \hat M_{add}) \hat \rho_s R_s} R_t.
     \label{eq:alphaTh}
\end{equation}
Experimental measurements of $\tan(\phi_d)$ as a function of $\Omega$, using same dataset as in Fig. \ref{fig:1}e,  are shown in Fig. \ref{fig:2}b for different string length. All those data points remarkably collapse on a single curve $\tan(\phi)=\alpha_{exp} R_t$ when plotted as a function of the trajectory radius $R_t$ (Fig. \ref{fig:2}c). This confirms the scaling law predicted in Eq. \ref{eq:phiD}, and the measured prefactor is $\alpha_{exp}=0.57$ cm$^{-1}$. Taking the specific values from our experiment $\hat \rho_s = 0.9$, $\hat S_{im} \approx 0.9$ (see SM), $\hat M_{add} = 1/2$ and $C_d \approx 0.4$, the prediction from Eq. (\ref{eq:alphaTh}) gives $\alpha_{th}= 0.17$ cm$^{-1}$. Although we used some approximation to evaluate the prefactor, the important discrepancy between theory and experiments suggests that other forces may have been forgotten in the model. 

\section{Lift force}

A good candidate of such force is the lift produced by the fluid. The latter arises from the complete solid rotation performed by the sphere during its revolution (see Fig. \ref{fig:2}a) \cite{falkovich_fluid_2018,rubinow_transverse_1961} and results from asymmetric flow between the two sides of the sphere. The flow can be vizualized by seeding the liquid surface with small particles (pepper grains) while water is made opaque using a bit of powder milk to ensure good contrast. A camera placed above the tank recorded 150 frames/s during a full revolution of the ball, ensuring small displacement of the ball and the particles between two images. A stack of 15 consecutive images is shown in Fig. \ref{fig:2}d. The same stack in the co-moving frame shows the streamlines bending by the ball. From this picture, one can qualitatively expect a lift force directed toward the inward of the trajectory as drawn in Fig. \ref{fig:2}d.

There exists exact formula to compute the lift force at low Reynolds numbers \cite{rubinow_transverse_1961} or in the case or irrotational flow through Kutta-Jukowski theorem \cite{falkovich_fluid_2018}. In our case, the Reynolds number is large but not infinite ($Re\sim10^3$) and the flow is strongly rotational on an area comparable to the ball's surface, as shown by the large vortex structure of Fig. \ref{fig:2}d. An analytical computation of the lift force is thus difficult to perform, but we nevertheless estimate its magnitude through experimental measurement of the flow. We reconstruct the velocity field at the water surface by particle image velocimetry (PIV) with the free software PIVLab \citep{thielicke_pivlab_2014,thielicke_particle_2021}
A typical result of velocity reconstruction in the comoving frame is shown in Fig. \ref{fig:2}e with the norm of the velocity superimposed in the background. The fluid's velocity is higher in the region closer to the rotation center (inner part) compared to the external region. We estimate the mean inner and outer velocity $V_{in}$ and $V_{out}$ by averaging over the corresponding areas (see Fig. \ref{fig:2}e). Both values are very sensitive to the frame change that is performed numerically, but their difference $\Delta V = V_{in}-V_{out}$ is insensitive to it. It is plotted as a function of the rotation speed $\Omega$ in Fig. \ref{fig:2}f and is compatible with a linear fit $\Delta V = 2 R_{eff} \Omega$ with $R_{eff} = 0.45$ cm. The errorbars are obtained by computing the standard deviation over $20$ consecutive PIV-frames and show low dispersion of the measurement along time. 

If one assumes non-slip boundary condition of the fluid at the ball's edges, the fluid in the comoving frame has velocity $V_{ball} + \delta \Omega$ on the inside and $V_{ball} - \delta \Omega$ on the outside, with $\delta$ the horizontal radius of the sphere at the surface (inset of Fig. \ref{fig:2}f). The velocity difference is thus $\Delta V = 2 \delta \Omega$. Remarkably, the found value $R_{eff}=0.45$ cm is very close from the horizontal radius of the sphere at the surface $\delta=0.5$ cm. The pressure difference between the two sides of the ball can be estimated from Bernouilli's equation $\Delta P = \frac{1}{2}\rho(V_{int}^2-V_{ext}^2) \approx \rho V_{ball} \Delta V$, where we neglected the $\Delta V^2$ term since $\Delta V / V_{ball} = \delta/R_t <0.2$. Keeping the scaling $\Delta V \sim 2\Omega R_s$, we can estimate the Magnus force over the whole sphere as
\begin{equation}
    F_L \approx \Delta P S_{im}  \approx M_L R_t \Omega^2 
    \label{eq:fMagnus}
\end{equation}
where $M_L \sim \rho \pi \hat S_{im} R_s^3$ is homogeneous to a mass and depends on the mass and radius of the ball. 

The lift force exactly scales as the intertial force but is directed in the opposite direction. Adding it to the previous force analysis leads to
\begin{equation}
    \tan(\phi)= \frac{F_v}{F_{in}-F_L} = \frac{\alpha_{th}}{1-M_L/(m+m_{add})} R_t
    \label{eq:phiM}
\end{equation}
and the ratio at the botton is the ratio of the lift force over the intertial force. The linearly measured relationship $\Delta V \propto\Omega$ ensures that Eq. (\ref{eq:phiD}) and Eq. (\ref{eq:phiM}) have similar dependency with respect to $R_t$, but with an increasing prefactor in the second case. This is perfectly consistent with the results of Fig. \ref{fig:2}c. Moreover, the order of magnitude of the lift mass is $M_L \sim \rho R_s^3 \sim m$, so that the denominator can be significantly impacted it. The found value $\alpha_{exp}=0.57$ cm$^{-1}$ found above prescribes $M_L/(m+m_{add}) = 0.7$. Thanks to the deformable string, we were thus able to measure the ratio of the lift force over the inertial force, or equivalently the ratio of the lift mass over the added mass.

Assuming as before that $m_{add} = m/2$, we get $M_L =1.05 \times m$, which is in fair agreement with the scaling law found above that gives $M_L \sim 0.75 m$. Our angle and velocity measurement thus offer two independent and consistent estimate of the lift's magnitude. A full quantitative measurement of the lift force would require to obtain the full pressure and velocity field around the sphere, as well as to take into account possible viscous effects occurring near the ball in the boundary layer. Such study is beyond the scope of this article, but it is interesting to compare our result with other lift force measurement. For three dimensional flows and high Reynolds numbers, the lift force acting on a rotating sphere of velocity $V$ is often set in the form $F_L = 1/2  C_L \rho S_{im} V_b^2$ with $C_L$ the lift coefficient that depends on spin factor $R_s \Omega/V_{b}$ \citep{nathan_effect_2008}. An approximate expression for the lift coefficient is $C_L \approx 2 C_D R_s \Omega / V_{ball}$ \cite{nathan_effect_2008}. We can thus write $F_L =   C_D \rho \hat S_{im} \pi R_s^3 R_t \Omega^2$, and we recover the same scaling law as in Eq. (\ref{eq:fMagnus}). The prefactor are moreover of the same order of magnitude since $C_D \sim 0.5$, but a quantitative comparison is not accessible despite our two independent measurement techniques. Indeed, the measurement from the angles requires an exact value of the added mass $m_{add}$ to extract the lift mass $M_L$, while a flow measurement over the whole sphere is not accessible for direct measurement of the lift. Moreover, the presence of an interface may affect the expression of the lift force compared to the fully immersed case. It is therefore not obvious whether the previous expression would hold or not beyond the order of magnitude that we have performed.

For completeness of the force analysis, we would finally like to mention that Eq. (\ref{eq:phiM}) predicts the dependency of the proportionality constant with respect to the characteristics of the ball, which essentially scales as $\alpha \sim 1/R_s$. A complete discussion and additional experimental data for different ball's radius were performed to probe the found scaling law. The results are presented in Supplementary Materials and are in fair agreement with the model.

\section{String deformation and effective model}
So far, we have shown that our experimental measurement were consistent with the force analysis that includes the lift force induced by the fluid. Nevertheless, this is not enough to predict the trajectories radius $R_t$ shown in Fig. \ref{fig:1}e as the force equilibrium admits an infinite number of solutions $(R_t, \phi)$. The geometry of the system will provide an extra relationship between those two quantities, which will finally fully determine the equilibrium state.

\begin{figure}
    \centering
    \includegraphics[width=\linewidth]{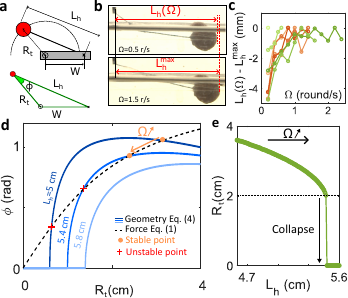}
    \caption{Apparent flexibility and effective model. (a) Triangle formed by the tensed string, the lever arm and trajectory radius. (b) Pictures from the side at $\Omega = 0.5$ round/s (top) and $\Omega = 1.5$ round/s (bottom) and sketch of the string elongation in the horizontal plane as $\Omega$ increases. (c) Experimental measurement of $L_h(\Omega) - L_h^{max}$ as a function of the rotation speed (same dataset as in Fig. \ref{fig:1}). (d) Angle $\phi$ as a function of the trajectory radius $R_t$ for different horizontal string length $L_h$ according to geometric constrain of Eq. (\ref{eq:phiG}) (color lines) and to force equilibrium of Eq. (\ref{eq:phiD}) (dashed line). When solutions exists, only the larges one (orange circle) is  stable (see supplementary materials). Due to varying horizontal elongation, the equilibrium radius changes with $\Omega$.  (e) Radius of the equilibrium trajectories as a function of $L_h$. The loss of solution for largest $L_h$ results in the collapse of the trajectory.}
    \label{fig:4}
\end{figure}

In the case where the strand is shaped as a straight line in the horizontal plane, the triangle formed by the ball, the hooking point and the rotation center in the horizontal planes (Fig. \ref{fig:4}a) imposes 
\begin{equation}
    \phi = \arccos\left(\frac{L_h(\Omega)^2+R_t^2-W^2}{2 L_s R_t} \right)
    \label{eq:phiG}
\end{equation}
where $L_h(\Omega)$ in the length of the string in the horizontal plane. The latter was found experimentally to increase with the rotation rate, as shown by experimental pictures taken from the side in Fig. \ref{fig:4}b. The horizontal projection $L_h(\Omega)$ reaches its maximal value $L_h^{max}$ when the string is fully tensed as in Fig. \ref{fig:4}b for $\Omega = 1.5$ round/s. We plot in Fig. \ref{fig:4}c the difference $L_s(\Omega)-L_h^{max}$ measured experimentally on the images taken from below. The maximum difference of about 4 mm for the smallest rotation speeds, and the horizontal length reaches its maximal value for $\Omega$ typically larger than $1$ round/s. Heuristically, this can be understood by noticing that the tension in the string increases with $\Omega$ as a reaction for the increasing viscous drag on the sphere $F_v \propto \Omega^2$., The string thus becomes straighter, which leads to an increase of the horizontal length $L_h(\Omega)$. A quantitative prediction of the string elongation as a function of the rotation speed requires a precise description of the mechanical properties of the string as well as of its interaction with the liquid, which is out of the scope of this work. For what follows, we will thus simply keep the progressive elongation of the string in the horizontal plane as a phenomenological fact.

The geometrical constrain (\ref{eq:phiG}) is plotted in Fig. \ref{fig:4}d for three horizontal length $L_h = 5, 5.4$ and $5.8$ cm (plain colored lines). We plot on the same graph the force equilibrium from Eq. (\ref{eq:phiM}) (black dashed line). The intersections between the two curves graphically determines the solutions ($R_t,\phi)$ that satisfy both force equilibrium and geometrical constrains for a given horizontal string length. Two solutions exist for $L_h =5$ cm and 5.4 cm, while no solution exists for $L_h = 5.8$ cm. Among the two solutions, we show in supplementary materials that only the largest one is linearly stable with respect to small radius perturbation. This solution is materialized by orange circles in Fig. \ref{fig:4}d, and corresponds to the radius effectively observed for a given horizontal string length. The latter increases with $\Omega$, which displaces the equilibrium radius $R_t$ as shown in Fig. \ref{fig:4}d (orange arrow). In particular, one expects the radius to decrease as the rotation speed increases, which is consistent with our experimental results. 
 
Such decreasing behavior is confirmed in Fig. \ref{fig:4}e, where we plot the expected radius $R_t$ a function of the horizontal length $L_h$. When the string's horizontal length is larger than a critical value $L_h^c \approx 5.5$ cm, no more stable solutions can be found (Fig. \ref{fig:4}c-d). This is in good agreement with our experimental results of Fig. \ref{fig:1}e, in which collapsed states occur for $L_s > 5$ cm.  In this case, the ball's equilibrium and the string's tension cannot be guaranteed simultaneously, which leads to a shape change of the strand. The critical horizontal length is reached for smaller rotation rates when the total length of the string $L_s$ is greater. The model therefore predicts that longer strings should collapse at lower rotation rates. Moreover, the smallest radius observed before collapse is approximately 2 cm and independent of $L_h$ (dashed line in Fig. \ref{fig:4}d). All those features are in quantitative agreement with our experimental results of Fig. \ref{fig:1}e.

Our model also sheds light on the existence of diverged states for shorter string length. The elongation eventually saturates at large rotation speed, so that a collapsed state cannot be reached if the string is too short $L_h^{max} < L_h^c$. At the bifurcation toward diverging state, we observed that the immersed volume of the ball suddenly decreases (see SM). This is consistent with our force analysis and Eq. (\ref{eq:phiD}): a decrease in the immersed volume decreases the angle $\phi$, which corresponds to increase the rotation radius $R_t$ as the rope aligns with $\vec e_r$ (see Fig. \ref{fig:2}a). Similarly, a decrease of the immersed volume tends to bring the dashed curve in Fig. \ref{fig:4}b closer from zero, which tends to increase the equilibrium radius. The reason for such brutal variation of the immersed volume, at the heart of the sudden radius jump, is nevertheless still to be understood. 

We have shown that the combination of force analysis on the ball and changing geometry of the string offers a consistent description of our experimental results. The apparent flexibility of the string in the horizontal plane, induced by its three-dimensional spatial reorganization, is at the heart of all the observed results. We therefore anticipate that the transition between rotating and collapsed trajectories described here is a rather general process, while the thresholds for their appearance depends on the chosen experimental configuration. We have already observed qualitatively the same three 'rotating', 'collapsing' and 'diverging' regimes for different ball's size and rotation arm's length, but a more systematic study would be required  to probe quantitative agreement with our model.

\section{Hysteresis and self-trapping}
 
\begin{figure}
    \centering
    \includegraphics[width=8.5cm]{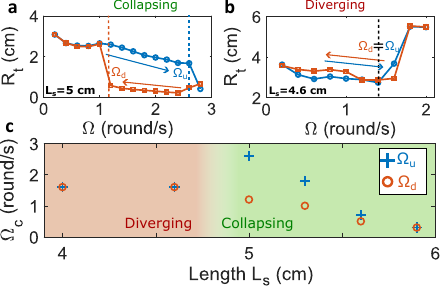}
    \caption{Hysteresis and critical rotation speed. (b) Radius of the trajectory for the diverging  (left, $L_s = 4.6$ cm) and collapsing (right, $L_s = 5.0$ cm) cases when the rotation speed is progressively increased (blue circles) or decreased (red squares) (c) Critical rotation speeds $\Omega_u$ and $\Omega_d$ as a function of string length $L_s$, showing the range of rotation speed for which the hysteresis exists.}
    \label{fig:5}
\end{figure}

We now aim to discuss the stability of the collapsed state. Starting from the collapsed state obtained for $L_s = 5$ cm and $\Omega > \Omega_u = 2.6$ round/s (blue circles in Fig. \ref{fig:4}a), the rotation speed is progressively decreased (red squares). Interestingly, the ball does not reach its previous rotating state but rather remains trapped at the rotation center. Such bistability occurs on a large range of rotation speed and eventually breaks down for $\Omega < \Omega_d = 1.2$ round/s where circular trajectories appear again. The rotation range $[\Omega_d, \Omega_u]$ for which the hysteresis occurs strongly depends on the string length and rapidly decreases as the latter increases, as shown in Fig. \ref{fig:4}c. In comparison, no hysteresis was observed in the diverged cases as shown in Fig. \ref{fig:4}b.

\begin{figure*}
    \centering
    \includegraphics[width=17cm]{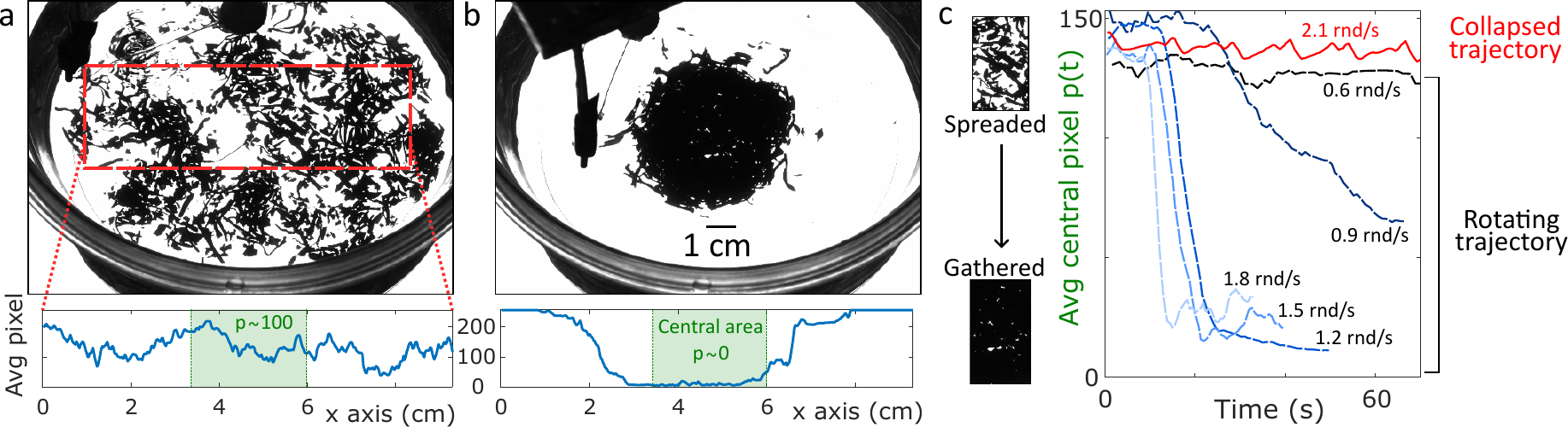}
    \caption{Tea-leaf paradox with a deformable stirrer. (a) Top view of the experimental setup, with tea-leaves uniformly spread at the bottom. Their spatial distribution is probed by computing the average in the vertical direction of the pixel's value in the red rectangle. (b) Occurence of tea-leaf paradox for $\Omega=1.2$ round/s and $L_s=5.5$ cm. The spatial distribution exhibits two strong plateau corresponding to full-black or full-white area. (c) Average pixel's value in the central area along time, showing or not the transition for uniform spreading to gathering of tea-leaves.}
    \label{fig:6}
\end{figure*}

The existence of hysteresis and memory effects is ubiquitous in fluid mechanics, and occurs for instance in Basset forces on accelerating objects \cite{michaelides_hydrodynamic_2003}, in the motion of bouncing droplets \cite{couder_walking_2005} or more generally in subcritical instabilities \cite{douady_experimental_1990,godreche_pattern_1998}. In the present case, the memory can be qualitatively interpreted in two ways. First, the tension lost in the string needs to be recovered to expel the ball from the flow's center. Second, the surrounding flow exhibits circular symmetry that needs to be broken to recover circular trajectories. Both features require a strong perturbation to be broken, leading to a robust hysteresis. 

Contrary to the diverging/collapsing phenomena, which was observed in all the tested configurations, the occurrence and range of hysteresis strongly depends on the peculiar experimental details. For instance, no hysteresis was observed when the guiding tube was set 1.5 cm above the water surface. In this case, the collapsed state presents some intermittency instead of regular spinning. The ball alternates between no motion, during which the string wraps around it, and sudden fast rotation where of the tension accumulated in the string releases. Those 'stop-and-go' trajectory provoke a lost of hysteresis, as the ball is sufficiently perturbed to reach back its rotating state. This effect is believed to come from the vertical component of the string's tension. It is consistent with the large hysteresis observed when the guiding tube is brought very close (3 mm) from the water surface to ensure horizontal pulling. We show in Supplementary Materials that in this case, very large hysteresis cycles could be observed with $\Omega_u = 2$ round/s while $\Omega_d < 0.3$ round/s.

\section{Tea-leaf paradox with a deformable spoon}

We discuss in this last section the secondary 'tea-leaf' flows induced by the motion of the sphere. To get closer from the original version of the tea-leaf paradox occuring in a cup, we used a circular tank of 13 cm diameter filled with 4.5 cm of water. We introduce infused tea-leaves in the water, which sink typically within a couple of seconds before settling at the bottom of the tank. A camera is placed above the tank while a LED panel is set below, so that the tea-leaves appear as dark areas (see Fig. \ref{fig:5}a-b). This allows to monitor their spatial repartition by simple contrast analysis. In each experiment, the liquid is initially at rest and the tea leaves are uniformly spread at the bottom. We then start the ball's rotation at $\Omega$ and record a movie from the top during one minute. The pictures in Fig. \ref{fig:5}a-b show the initial and final state for $\Omega = 1.2$ round/s and $L_s = 5.5$ cm. The leaves clearly gathered at the rotation's center, showing that the tea-leaf paradox also occurs with our 'deformable spoon'. As mentioned in the introduction, the migration is caused by the secondary flow induced by the boundary layers between the rotating bulk and the non-moving edges of the tank \cite{einstein_ursache_1926}. Our results show that adding some flexibility on the stirrer does not qualitatively impact the flow generation.

However, the deformation of the spoon can strongly impact the efficiency of those secondary flow and so particle's transport. In particular, the leave's transport is suppressed when the ball reaches a collapsed state. To prove this, we first compute the leave's spatial distribution at a given time along the horizontal direction by performing the average of the pixel's value over a small vertical window, as shown in Fig. \ref{fig:5}a. The resulting curves display either uniform (Fig. \ref{fig:5}a) or plateau-like (Fig. \ref{fig:5}b) distribution, corresponding respectively to uniformly spread or gathered tea-leaves. Those two configurations can be easily distinguished by computing the average value $p$ on the central area materialized by the green area in Fig. \ref{fig:6}a-b. Uniform spreading corresponds to $p\sim100-150$ (equal mix of white and black pixels) while gathered leaves correspond to $p\sim0$ (completely black pixels). The value of $p(t)$ along time is thus a good indicator of the leave's transport due to secondary flow. It is plotted in Fig. \ref{fig:5}c for different rotation speeds. For the lower rotation speed $\Omega=0.6$ round/s, $p(t)$ is approximately constant which indicates that no transport occurred during the acquisition. This suggests that the secondary flows are too weak to transport the leaves. For higher rotation speeds $\Omega = 0.9-1.8$ round/s, the ball performs circular trajectories and we clearly observe the gathering of the leaves after some delay. Interestingly, this delay seems to decrease as the rotation speed increases, which shows an increase in the mixing efficiency with the rotation speed. The same trend was observed for another dataset generated for $L_s = 4.5$ cm, but more data would be needed to perform statistical analysis.

Interestingly,  a further increase the rotation speed to $\Omega=2.1$ rad/s completely suppresses the tea-leaf transport. This loss of the tea-leaf paradox is concomitant with the appearance of a collapsed state, leading to sudden decrease of the mixing area. We have let the system evolve in this state for tens of minutes without observing any noticeable migration of the leaves. This suggests that the secondary flow are, as for the lowest rotation case, too weak to perform transport. Those results clearly show that in the presence of a flexible stirrer, increasing the rotation speed does not always lead to higher mixing efficiency. A sudden drop of the mixing area can indeed occur when the stirrer reaches a collapsed state, which significantly reduces the momentum transfer between the fluid and the sphere.

\section{Conclusion}
We have discussed in this work the circular motion of a ball pulled by a string and the induced flows. The unexpected radius shrinking when the rotation speed increases results from an equilibrium between the string's deformation and flow's action. Our deformable system allowed to measure the ratio between the lift and the added mass for a sphere moving circularly at the water surface, and the obtained results were consistent with direct flow measurement. At higher rotation speed, we have shown the existence of bifurcation and robust self-trapped states. Interestingly, those trapped state were shown do strongly decrease the mixing efficiency compared to rotating state at lower rotation speed.

The results presented in this work open several interesting avenues for the future. It would be interesting to perform the same experiment for other size or body shapes, as well as to study analytically the impact of the free surface on the lift force expression. It also strongly suggests to perform the same experiments with several objects to introduce pairwise interactions mediated by the flow, as it was done for rotating magnetic disks \cite{grzybowski_dynamic_2000}. Last, one may replace water with a non-Newtonian fluid, which were for instance shown to modify swimmers efficiency \cite{espinosa-garcia_fluid_2013}, or replace the inextensible rope by a flexible one to introduce flow-elasticity coupling. The impact of those modifications on the object's trajectories and induced mixing are expected to exhibit rich phenomenological features.

\section{Supplementary materials}
Supplementary materials of this article include details on the experimental setup, details on data selection for the force analysis, detailed computation on the stability of equilibrium position and additional experimental data with different balls.
\section{Acknowledgment}
The authors acknowledge J.B. Gorce for fruitfull discussions, M. Mallejac for his help with Blender and G. Noetinger for precious feedbacks.

\section{Data availability statement} All data and codes used for this study are available from the corresponding author upon reasonable request.
\bibliography{biblio}

\end{document}


\maketitle

\section{Experimental methods}
\subsection{Experimental details}
The wool strand that we used was $\sim 0.5$ mm thick. It is attached to the ball by drilling a small hole in one hemisphere, introducing the strand in it and filling the hole with glue. There is therefore no possible rotation between the string and the ball. Residual torsion in the strand is released by letting the ball freely hang in the air for a while before plunging it in water. The rotation arm was set 6.5 cm above the water surface, while the guiding tube is set $5$ mm above the surface.

A stepper motor (800Nmm, 2000min-1 1.8 $\deg$, NEMA 17) is commanded with an Arduino and a motor controller (BigEasy Driver) to rotate at a fixed rotation speed. We control that the obtained rotation speed was correct by looking at its rotation rate using the camera (Balser Ace). During an experiment, we let the ball straddle for 30s before measuring its trajectory during 10s. This ensures that the stationary regime is always reached, that the ball performs several turns on each movie and allows good repeatability of the results. We have tested to let the ball rotate during 2 minutes before performing the measurement. The difference with the one obtained obtained while letting the ball rotate for only 30 s was lying within the experimental uncertainty and could therefore not be distinguished.

The impact of the tank's edges is also believed to be irrelevant for the present experiment. First, the waves emitted by the sphere are typically of amplitude $0.2$ mm, which shows that their action on the ball can be neglected compared to the ball's inertia. On the other hand, the velocity of the fluid near the edges was always found to be negligible compared to the velocity of the ball or to the velocity of the fluid near the ball. This shows that potential recirculations or friction due to the edges can be discarded to study the ball's motion. Those secondary flow are nevertheless relevant for the tea leaf paradox mentioned at the end of the main text.

\subsection{Trajectory and angle measurement}
\begin{figure}
    \centering
    \includegraphics[width=10cm]{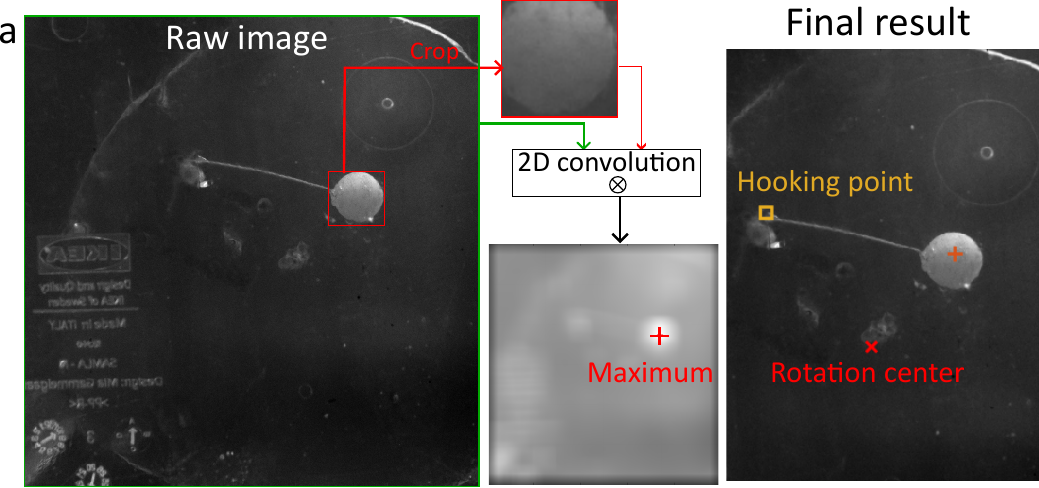}
    \caption{(a) Example of image analysis: a raw image (left) is 2D-convulted with the image of the ball (middle-top). The maximum of the resulting image (middle-bottom) has maximum value at the center of the ball. This allows to recover the ball's position (right). The position of the hooking point (yellow square) and of the rotating center (red X) is also shown. }
    \label{fig:S1}
\end{figure}

In order to recover the ball's trajectory, a camera is placed below the tank and records 10 images/revolution of the ball. A raw image is shown in Supplementary Fig. 1. The ball appears brighter than the background, but contrast analysis has shown some failure. We therefore performed 2D convolution of the raw image with the image of the ball, and the resulting image has clear maximum at the ball's location. Performing similar analysis on each image gives trajectories ($(x(t), y(t)$) as shown in Fig.1 of the main text. The rotation center (X cross in Supplementary Fig. 1) is obtained by computing the mean value of $x$ and $y$ over several revolutions. In order to measure the angle $\phi$, one also needs to get the position of the hooking point. The later is simply spot on a single image. We have checked that the particular image chosen does not affect significantly the measurement result.

\begin{figure}
    \centering
    \includegraphics[width=15cm]{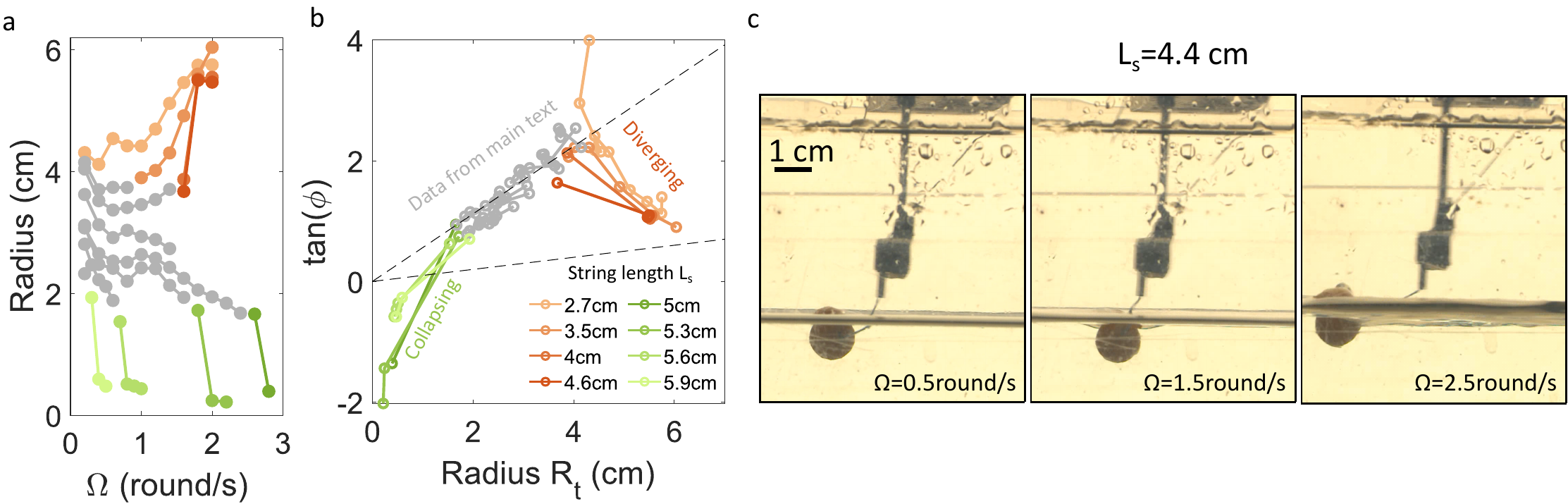}
    \caption{(a-b) (Colors) Excluded data from angle analysis in the main text, for which only the gray points were kept. (c) Side view of the ball for $L_s=4.4$ cm showing the up lift experienced by the ball for largest values of $\Omega$.}
    \label{fig:S2}
\end{figure}

\begin{table}[]
    \centering
    \begin{tabular}{ |c||c|c|c|c|c|c|c|c|} 
     \hline
     $L_s$ (cm) & 2.7 & 3.5 & 4 & 4.6 & 5 & 5.3 & 5.6 & 5.9 \\ 
     \hline
     $\Omega_{lim}$ (round/s) &  0 & 1 & 1.6 & 1.6 & 2.8 & 1.8 & 0.7 & 0.3 \\ 
     \hline
    \end{tabular}
    \caption{Maximum rotation value considered for each string length for data selection in angle measurement}
\end{table}

\subsection{Data selection for angle measurement}

As discussed in the main text, some data selection has been performed in the main text between radius measurement and angle measurement. First, all data corresponding to collapsed states have been discarded since the rope is not shaped as a straight line in those cases (green points in Supplementary Fig. 2). The proposed model was therefore irrelevant for those data, hence their exclusion.

We also decided to exclude all data associated with an increase of the radius, which corresponds to red-colored points in Supplementary Fig. 2 associated with $\Omega \geq \Omega_{lim}(L_s)$ with the value $\Omega_{lim}$ summarized in Supplementary Table 1. In particular, all data from $L_s=2.7$ cm were excluded. We indeed observed that when the radius increases, the immersed volume of the ball was diminishing. Such phenomena can be seen on Supplementary Fig. 2c, where the immersed volume of the ball decreases significantly between $\Omega=1.5$ round/s (non-diverging) to $\Omega=2.5$ round/s (diverging). As the proposed model assumes constant immersed volume, it is not surprising that the corresponding points do not match with the prediction (Supplementary Fig. 2b). Although it is beyond the scope of this work, this varying immersed volume could lead to other interesting regimes. In particular, our data suggest that the up-lift experienced by the ball appears for larger $\Omega$ and more brutally as the string length increases. The origin of this lift is not clear to us at this stage but is an interesting perspective of this work.

\section{Stability of equilibrium positions}

We have shown in the main text that the equilibrium angle $\phi$ must fulfill both the force equilibrium condition which reads
\begin{equation}
    \tan(\phi)= \alpha_{th} R_t
    \label{eq:phiDS}
\end{equation}
as well as the geometric condition deduced from the triangle formed by the ball, the hooking point and the rotation center
\begin{equation}
    \cos{(\phi)} = \frac{L_h^2+R_t^2-W^2}{2 L_s R_t}
    \label{eq:phiGS}
\end{equation}

A graphical analysis performed in the main text shows that this equation admits two solutions when $L_s$ is small enough and zero if it is too large. The latter case was interpreted as the collapsing behavior. When two solutions exists, we admitted in the main text that only the largest one is stable. This section is devoted to demonstrate this fact.

This can be shown by writing the equation of motion of the ball in the laboratory frame assuming constant rotation speed $\dot \theta = \Omega$ (which can be positive or negative) but not constant radius $r$
\begin{equation}
    (m+m_{add}) (\ddot r - r \Omega^2) \vec e_r + 2 (m+m_{add}) \dot r \Omega \vec e_\theta = \vec T + \vec F_v + \vec F_L
\end{equation}

\begin{figure}
    \centering
    \includegraphics[width=8cm]{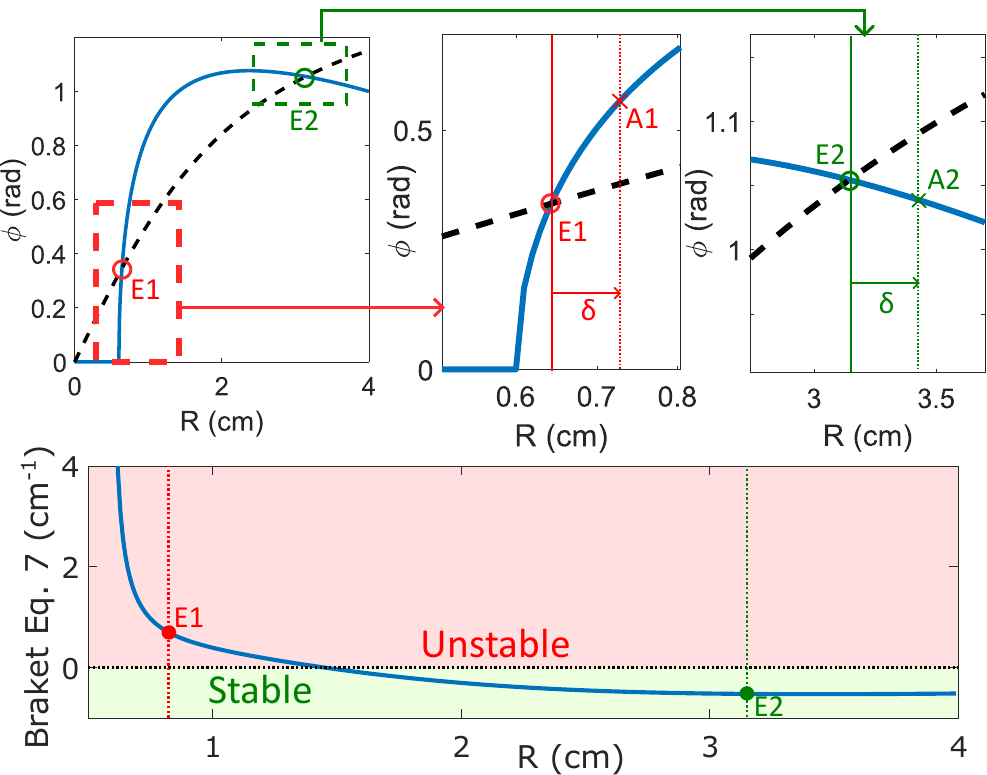}
    \caption{Stability analysis. Upper panels: angle $\phi$ determined from Eq. (\ref{eq:phiDS}) (black dashed line) and geometrical constrain (\ref{eq:phiGS}) (blue plain line). The two equilibrium positions are denoted E1 and E2, and insets show small perturbations A1 and A2 for stability analysis. The later is determined by the lower panel which plots $d \tan{\phi}/dR - 1/R_t$ from Eq. \ref{eq:deltaPoint}. Positive sign implies unstable equilibrium while negative sign  ensures stability.}
    \label{fig:enter-label}
\end{figure}

We will now perform some perturbation around the equilibrium radius $R_t$ and write $r=R_t + \delta$ with $\delta \ll R_t$. For simplicity, we will also assume that return at equilibrium occurs slowly so that $\dot \delta \ll R_t \Omega$. From this, we get to first order the viscous force as 
\begin{equation}
    \vec F_v = \frac{-1}{2}\rho C_D S_{im} R_t^2 \Omega \lvert \Omega \rvert \left(1+ \frac{\delta}{R_t}\right) \left(\frac{\dot \delta}{R_t \Omega} \vec e_r + (1+\frac{\delta}{R_t}) \vec e_\theta \right) 
\end{equation}
In order to keep things simple, we will consider that the lift force acts as an effective mass, so that we will perform the change $m_{add} \rightarrow m_{add} - m_L$ at the end to take it into account. After some calculation, one finds
\begin{equation}
    \tan{\phi} - \tan{\phi_{eq}} = \frac{ \delta}{R_t} - \frac{2 +  \tan{(\phi_{eq})}^2}{ \tan{(\phi_{eq})}} \frac{\dot \delta }{R_t \lvert \Omega \rvert} + \frac{\ddot \delta}{R_t \Omega^2}  
    \label{eq:tanPhi}
\end{equation}
with 
\begin{equation}
    \tan (\phi_{eq}) = \frac{F_v}{(m+m_{add}) R_t \Omega^2} = \frac{- \rho C_D S_{im} R_t}{2(m+m_{add})}\frac{\Omega}{\lvert \Omega \rvert} \approx \frac{3C_D \hat S_{im}}{8 (1 + \hat M_{add} - \hat S_{im} \hat M_L / \hat \rho_s) \hat \rho_s R_s} R_t
    \label{eq:alphaTh}
\end{equation}
where we used adimension all the quantities to emphasize the dependencies with the sphere's dimension. We therefore set $\hat \rho_s = m / (4/3 \pi R^3  \rho)$, $\hat M_{add} = m_{add}/m$, $\hat S_{im} = S_{im}/(\pi R_s^2)$ and $M_L =\hat M_L \rho \pi \hat S_{im} R_s^3$. Remember also that $\Omega <0$ in our convention.

We can now perform a linear stability analysis of the two equilibrium solutions which can be seen on Supplementary Fig. 3. To plot those graphs, we have taken the case $\Omega < 0$ resulting in $\tan \phi_{eq} > 0$ and so $\phi > 0$. For simplicity, we will neglect the $\ddot \delta$ term, which corresponds to assume that $\ddot \delta \ll \dot \delta \Omega$ consistent with the hypothesis of slow return to equilibrium. Eq. (\ref{eq:tanPhi}) can now be recast as
\begin{equation}
    \frac{\dot \delta}{V_0} = \left[ \left( \frac{d \tan{\phi}}{dR}  \right)_{\phi_{eq}} - \frac{1}{R_t} \right] \delta
    \label{eq:deltaPoint}
\end{equation}
with 
\begin{equation}
    V_0  = R_t \lvert \Omega \rvert \frac{\tan{(\phi_{eq})}}{(2 +  \tan{(\phi_{eq})}^2} >0
\end{equation}

We first consider the stability of the largest equilibrium position E2 (right panel of Supplementary Fig. 3). If one perturbs the equilibrium by pushing a bit the ball on the outside ($\delta >0)$ while keeping the string tensed so that one remains on the plain blue line, one has $\phi <\phi_{eq}$ (point A2 in the plot). Therefore, one has $ \frac{d \tan{\phi}}{dR}<0$ and the bracket in Eq. (\ref{eq:deltaPoint}) is negative, leading to $\dot \delta < 0$. The same analysis for $\delta < 0$ leads to $\dot \delta > 0$, and this equilibrium position is therefore stable.

On the other hand, one can consider the smallest equilibrium position E1 (middle panel of Supplementary Fig. 3) and a perturbation with $\delta > 0$ corresponding to A1. In this case, one has $ \frac{d \tan{\phi}}{dR}>0$ so that the sign in the braket of Eq. \ref{eq:deltaPoint} is undefined. Nevertheless, one can see on Fig. 3 that the derivative $d\phi/dR$ near E1 is very large ($\sim 3$ rad/cm) while $1/R \sim 1.5$ cm$^{-1}$. Therefore, the braket will be positive and one has $\dot \delta > 0$, so that the equilibrium position is unstable. This is confirmed by Fig. 3b which plots the braket of Eq. (\ref{eq:deltaPoint}) that changes sign between the two equilibrium points.

\section{Impact of the ball's size}
\begin{figure}
    \centering
    \includegraphics[width=\linewidth]{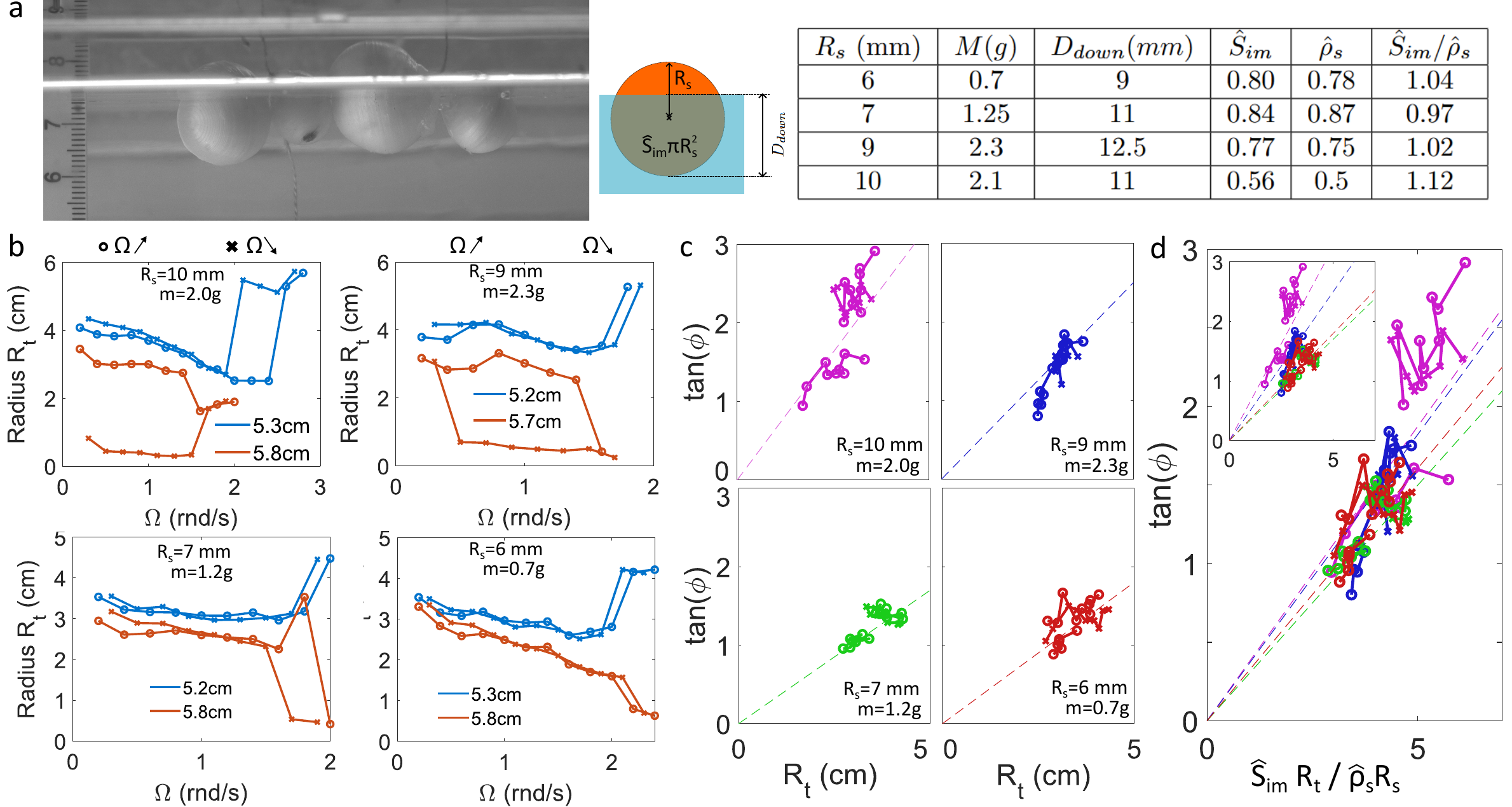}
    \caption{Impact of the ball's size on the force analysis. (a) Picture of the different ball used (the ruler is in cm), definition of useful quantities and summarizing table of the main characteristics of each ball. (b) Trajectory radius $R_t$ as a function of the rotation speed for four different balls of radius $R_s = 10, 9, 7$ and 6 mm and mass $m=2.0, 2.3, 1.3$ and 0.7 g. Circles correspond to progressively increasing $\Omega$ while crosses correspond to decreasing $\Omega$ to probe potential hysteresis effects (c) $\tan(\phi)$ as a function of trajectory's radius $R_t$ from the same dataset. (d) Same data gathered in the same plot with renormalized horizontal axis to check the validity of Eq. XX. Inset: same data as a function of non-renormalized $R_t$.}
    \label{fig:S4}
\end{figure}
In order to probe the generality of the previous results, we have performed the same experiment with four different balls of radius $R_s=6,7,9$ and $10$ mm and mass $m=0.7, 1.3, 2.3$ and $2.0$ g shown in Fig. \ref{fig:S4}a. The lever arm was chosen to be $L=5.1$ cm while the height between the guiding tube and the surface was 3 mm, a bit smaller than before. All the experiments were performed during the same day to ensure same experimental conditions. For all the spheres, we observed a decrease of the radius as the rotation speed increases, followed by a transition toward collapsed or diverged states depending on the string's length. This shows that the results described above are not limited to a narrow range of parameters but rather seem to be generic feature of the system. 

As in the previous experiment, we plot $\tan(\phi)$ as a function of the trajectory's radius $R_t$ in  Fig. \ref{fig:S4}b. In all cases, both quantities can be reasonably considered to be proportional to each other, those further confirming the scaling law predicted by Eq. (\ref{eq:alphaTh}). The slope's coefficient $\alpha_{exp}$ is found to strongly depend on the considered ball and goes from $\alpha_{exp}=0.65$ cm$^{-1}$ for $R_s = 6$ mm to $\alpha_{exp}=0.35$ cm$^{-1}$ for $R_s = 10$ mm. The Eq. (\ref{eq:alphaTh}) predicts the variation of the prefactor as a function of the peculiar properties of the ball $\hat S_{im}, \hat \rho_s$ and $R_s$, which are gathered in Fig.\ref{fig:S4}a. The ratio $\hat S_{im} / \hat \rho_s$ is approximately constant and unitary for all the spheres. For simplicity, we will therefore probe a slightly simpler scaling law than the one from Eq. (\ref{eq:alphaTh}) and take
\begin{equation}
    \tan (\phi_{eq}) \approx \frac{3C_D }{8 (1 + \hat M_{add} - \hat M_L )} \frac{\hat S_{im} R_t}{\hat \rho_s R_s} 
    \label{eq:alphaTh}
\end{equation}
We test the precited scaling law by plotting in Fig. \ref{fig:S4}c the same $\tan{\phi}$ as a function of the adimensioned trajectory radius $\hat S_{im} R_t / (\hat \rho_s R_s$. Such rescaling reduces the spreading of data compared to the non-renormalized case (inset of Fig. \ref{fig:S4}c). In particular, the slope coefficient $\alpha_{exp}$ now lies between $[0.30, 0.38]$, compared to $[0.34, 0.65]$ cm$^{-1}$ before rescaling. All the rescaled data can be reasonably fitted by a linear law with slope $\alpha \approx 0.35$. This finally gives 
\begin{equation}
    \hat M_L \approx 1 + \hat M_{add} - \frac{3C_D}{8\alpha}\approx \hat M_{add} + 0.6
\end{equation}

